\documentclass[12pt]{article}
\usepackage[american]{babel}
\usepackage[utf8]{inputenc}   
\usepackage[T1]{fontenc}      
\usepackage{csquotes}
\usepackage{microtype}
\usepackage[round, authoryear]{natbib}
\usepackage{hyperref}
\usepackage{graphicx}
\usepackage{setspace}
\usepackage{geometry}
\usepackage{tabularx, array}
\title{\vspace{2cm}\textbf{Report on the Scoping Workshop on AI in Science Education Research}\vspace{1cm}}
\author{}
\date{}

\begin{document}

\begin{titlepage}
    \centering
    {\Large \textbf{Report on the Scoping Workshop on AI in Science Education Research}\par}
    \vspace{2cm}
    
    {\large \textbf{Organizers:}\\
    Marcus Kubsch, Marit Kastaun, Peter Wulff, Nicole Graulich\par}
    \vspace{1cm}
    
    {\large \textbf{Participants:}\\Moriah Ariely, Alexander Bergmann-Gering, Sebastian Gombert, Bor Gregorcic, Hendrik Härtig, Benedikt Heuckmann,  Andrea Horbach, Christina Krist, Gerd Kortemeyer, Ben Münch, Samuel Pazicni, Joshua M. Rosenberg, Sascha Schanze, Gena Sbeglia, Vidar Skogvoll, Christophe Speroni, Christoph Thyssen, Lars-Jochen Thoms, Brandon J. Yik, Xiaoming Zhai\par}
    \vspace{2cm}
    
    {\large \today\par}
\end{titlepage}

\tableofcontents
\newpage
\section{The Goals of This Document}

The purpose of this report is to provide a structured overview of how artificial intelligence (AI) is reshaping science education and to outline the implications for the science education research community. Specifically, the document seeks to (a) clarify definitions of AI in order to avoid conceptual ambiguity, (b) synthesize current evidence of AI’s influence on scientific practice, science teaching, and science education research, (c) map emerging areas of inquiry where AI introduces new opportunities and challenges, and (d) identify methodological, ethical, and infrastructural implications that should guide future work. In doing so, the paper aims to establish a common foundation for dialog between stakeholders, foster a principled and responsible integration of AI into science education, and inform future research, policy, and practice in this rapidly evolving field. The report is the result of a 2-day expert scoping workshop on AI in science education research funded by the VW Foundation. Statements that are not supported by citations reflect perspectives and insights from participants. 

\section{Artificial Intelligence in Science Education Research}
We want to start this section with a note on terminology. In science education research, \textit{artificial intelligence} (AI) functions as an umbrella term that obscures the vivid distinctions among its varied subfields. A broad definition such as “computers which perform cognitive tasks […] an umbrella term to describe a range of technologies and methods, such as machine learning (ML), natural language processing (NLP), data mining, neural networks or an algorithm” effectively captures this diversity \citep{usdepartment2023}. To avoid ambiguity, researchers should explicitly specify which AI “flavor” they engage \citep{narayanan2024aisnakeoil} (for instance, machine learning (algorithms that learn from data and generalize to new cases) \citep{bishop2006}, deep learning (a specialized branch of ML using multi-layer neural networks for tasks such as vision and language processing) \citep{goodfellow2016}, or generative AI (models that generate new content based on learned patterns) \citep{brown2020}) and how they implement it in software unless they actually refer to AI at large so that the umbrella term is indeed justified. 

In educational contexts, failing to clarify these distinctions risks conflating mechanisms, underestimating biases, and overgeneralizing pedagogical implications \citep{zawacki2019}. For example, rule-based expert systems operate through explicit knowledge representation and logical inference, while deep learning networks extract patterns through multilayered transformations of input data. This distinction has profound implications for conceptualizing intelligence, reasoning, and knowledge construction \citep{vanRooij2024reclaiming}. By precisely naming the sub‐domain—ML, deep learning, NLP, expert systems, etc.—researchers can support clarity in methodology, align theoretical frameworks, illuminate systems' affordances or constraints, and responsibly address ethical considerations. Such specificity is a prerequisite for robust, interpretable, and actionable insights into how particular AI methodologies influence learning, assessment, and instructional design. 

\subsection{AI is Changing Science}

In the physical sciences, AI (mostly deep learning) has enabled major breakthroughs in pattern recognition and data processing. At CERN’s Large Hadron Collider, machine learning filters vast collision datasets to detect rare events such as the Higgs boson, tasks impossible for traditional methods \citep{radovic2018}. Astronomical research similarly leverages convolutional neural networks to accelerate analysis of gravitational lensing \citep{hezaveh2017}. In the life sciences, AlphaFold achieved unprecedented accuracy in protein structure prediction, solving a problem that had remained intractable for decades \citep{jumper2021}. DeepMind’s GNoME accelerated crystallographic discovery by centuries, compressing vast knowledge into computationally feasible timeframes. AI may also support precision medicine by integrating multi-modal patient data to improve diagnosis and treatment design \citep{Boehm2022MultimodalPrecisionOncology}. Climate and Earth sciences have embraced foundational AI models like NASA’s Prithvi, which significantly improve local and regional weather forecasts \citep{reichstein2019}. Employing AI, Diffenbaugh and Barnes \citep{diffenbaugh2024} have shown that global warming is likely to surpass 2°C even if current emission reduction targets are met, underscoring AI’s power to refine predictive models. 

AI’s increasing autonomy raises new epistemological issues. Systems such as Coscientist can autonomously design and execute chemical experiments \citep{boiko2023autonomous}, while NASA’s Perseverance rover uses AI to navigate Mars and select sampling sites independently \citep{gaines2016}. These cases exemplify AI's productive functioning as a scientific agent rather than a mere instrument. Yet this autonomy introduces interpretability challenges. In physics, predictions can often be benchmarked against established theories, but in medicine, opaque “black box” outputs raise concerns \citep{rudin2019, pdg2024}. Scholars emphasize epistemic vigilance, supported by heuristics such as the AIR model—Aims and values, Ideals, Reliable processes—to evaluate AI-generated claims \citep{chinn2014epistemic}. Failures in reproducibility, particularly in biomedical AI research, highlight the urgency of robust standards \citep{mcdermott2025meds}. Initiatives like REFORMS set benchmarks for responsible ML use across domains \citep{kapoor2024}. 

AI’s benefits come with environmental costs. Data centers consumed 460 terawatt-hours in 2022, placing them among the world’s largest electricity consumers \citep{iea2024}. They also generate electronic waste, consume water, and rely on environmentally damaging mining of rare minerals. 

\subsection{AI is Changing Science Education}

AI is poised to reshaping science education by transforming instructional practices, learning environments, and educators’ roles \citep{UNESCO2023_GuidanceGenAI}. A recent systematic review spanning empirical studies from 2014 to 2023 highlights AI's multifaceted impact: enhancing instructional design, adapting assessments, and influencing both student and teacher experiences in science classrooms \citep{almasri2024}. In particular, the integration of AI-driven tools in science teaching has produced measurable improvements in learning outcomes \citep{kestin2025}.  At the same time, integrating AI into science education entails the risk of inversion effects—for example, over reliance that reduces students’ cognitive engagement (Bauer et al., 2025). Researchers therefore call for systematic, evidence-based approaches to AI integration.

Innovative AI applications are facilitating new forms of collaborative learning. For instance, the CLAIS prototype (Collaborative Learning with AI Speakers) positions AI as a peer within group science-learning activities. Pre-service elementary science teachers who engaged with CLAIS reported increased pedagogical and technological content knowledge, and recognized changes in epistemic practices—suggesting AI’s potential to reshape classroom discourse and co-construct knowledge \citep{lee2023}. 

Generative AI is opening pathways for rapid and customized content creation. Researchers have demonstrated that AI can produce physics simulations via generative models, enabling the development of virtual labs on-the-fly. These AI-generated simulations offer dynamic, personalized explorations of physical phenomena, laying the groundwork for content tailored to diverse learners \citep{benzion2024}.

Teacher attitudes toward AI in science education are generally positive. A study examining science teachers’ acceptance of AI found high levels of perceived self-efficacy and expectations of ease of use in AI-supported classrooms \citep{aldarayseh2023, esbenshade2025emergingpatternsgenaiuse}. Pilot interventions embedding AI directly into science curricula point at the importance of integrating AI concepts within disciplinary contexts, rather than teaching AI as an abstract, standalone topic. One study piloted AI-related lesson packages in science classrooms, helping students see AI’s relevance to scientific practices and fostering more meaningful connections \citep{park2023}. AI can also assist in the design of entire curricula, such as for complex systems in physics \citep{crokidakis2023}.

\subsection{AI is Changing Science Education Research}

Artificial intelligence has begun to reshape science education research in ways that extend beyond introducing a new topic of study—it is transforming how knowledge is produced, validated, and disseminated within the field. A systematic review of 36 AI-related contributions at the National Association for Research in Science Teaching (NARST) 2024 conference shows that AI is currently used as both a methodological and thematic catalyst. It functions as a multifunctional, multimodal, and generative tool, while simultaneously raising new ethical issues concerning learning goals, curriculum content, individualized feedback, teacher education, bias, and equity \citep{lee2025}. 

From a methodological perspective, AI is altering the conduct of research itself. AI-based systems are increasingly used to support peer review—for example, through automated checks of language, data, and references, or by suggesting improvements to manuscripts \citep{naddaf2025}. This evolution reshapes quality assurance in scientific publishing and creates new demands for transparency and governance. Within science education research, multimodal large language models (MLLMs), which integrate text, image, and audio, are enabling new forms of empirical inquiry and feedback scenarios that are more personalized, interactive, and accessible. Such systems support not only students but also teachers in developing tailored, adaptive instructional formats \citep{kuchemann2024large}. 

Conceptually, these developments prompt researchers to revisit foundational questions about interpretation, validity, and the role of human judgment. When algorithms participate in data analysis, feedback generation, or even hypothesis formation, the boundary between human and computational cognition becomes a productive site of inquiry. This raises pressing questions about construct validity, transparency, and the ethics of algorithmic mediation in research.

These shifts directly motivate the \textit{areas of inquiry} outlined in Section~\ref{sec:areas}, where AI now intersects with virtually all dimensions of science education research—from curriculum and assessment to epistemology, inclusivity, and teacher learning. As AI systems demonstrate impressive but brittle ability on STEM problem-solving tasks, researchers must ask which forms of reasoning, creativity, and metacognition remain uniquely human and educationally valuable. Likewise, the expanding role of AI in teacher professional development and student inquiry invites renewed attention to issues of agency, trust, and expertise. 

At the same time, AI transforms the \textit{ways of inquiry} discussed in Section~\ref{sec:ways}. Quantitative researchers increasingly combine classical inference with model-based discovery; qualitative analysts use AI-assisted coding to scale interpretation; and design-based or ethnographic approaches integrate AI as both a methodological instrument and a participant within learning environments. These developments signal the emergence of a ``cyborg methodology,'' where human interpretive labor and algorithmic assistance are deliberately combined to expand what can be seen, measured, and theorized in science education research. 

Overall, AI does not replace the epistemic and methodological foundations of science education research—it extends them. It broadens what can be observed, enhances methodological reflexivity, and challenges researchers to reconsider how evidence, theory, and ethics coevolve in an AI-mediated research landscape \citep{lee2025,kuchemann2024large}.

\section{Areas of Inquiry in Science Education Research and How AI is Affecting Them}
\label{sec:areas}

Science education research encompasses multiple strands, ranging from how students learn disciplinary content to how teachers are prepared, how curricula and assessments are designed, and how social, cultural, and ethical contexts shape learning. Canonical maps of the field, such as NARST’s strands and historical syntheses in physics education research, provide a scaffold for situating current AI-related questions within longer-standing agendas \citep{narst_strands,mcdermott1999,cummings2011}. Before the advent of contemporary AI tools, science education research had already identified enduring challenges: supporting students in developing epistemological understanding and critical thinking as foundations for meaningful learning \citep{madsen2015}, addressing the persistent misalignment between assessments and authentic scientific practices, and confronting concerns about equity and teachers’ use of assessment data \citep{liu2016,zhai2021}. AI now intersects with, accelerates, and complicates this landscape.

\subsection{Baseline: Performance of AI Systems on STEM Tasks}
Understanding what AI systems can (and cannot) do on core STEM tasks is a prerequisite for principled research design and policy. Studies probing performance on multimodal tasks—such as interpretation and construction of graphs or symbolic mathematics—establish baseline capability and failure modes \citep{polverini2024a,kortemeyer2025,purandare2025}. These baselines inform where to deploy AI productively (e.g., feedback on representational competence) and where human scaffolding remains essential.

\subsection{Curriculum Development and Disciplinary Learning}
If general-purpose models can already solve substantial portions of undergraduate STEM problem sets, curricula must pivot toward goals that privilege modeling, explanation, multiple representations, and the quality of reasoning over final answers \citep{pimbblet2025,tschisgale2025}. In chemistry and physics, scholars outline concrete implications for curricular content and AI-integrated tasks that mirror authentic scientific practice \citep{berber2025,polverini2024b}. Generative AI also supports design-based research on learning—for instance, as tutoring or simulation engines that personalize complex-systems explorations and computational modeling \citep{kestin2025,chabay2008,caballero2012,perezlinde2025,yeadon2024}. These affordances raise perennial questions in new guises: Which tasks remain instructionally valuable when AI can draft code or write lab reports? What does productive struggle look like when routine work is automated?

\subsection{Assessment and Feedback}
A rapidly growing line of work investigates AI for formative feedback, grading support, and exam design, emphasizing derivations and reasoning, with mixed but promising evidence \citep{wilson2024,chen2025,wan2024,liu2016,wang2025}. Studies have also explored diagnosing misconceptions and supporting adaptive instruction \citep{kokver2025artificial}. At the same time, equity and bias remain central: language proficiency, handwriting, perceived gender/ethnicity, and prior knowledge can interact with AI pipelines in complex ways \citep{wang2025,zhai2021}. Work on academic integrity documents AI-enabled cheating and its impacts, motivating new assessment designs and integrity supports \citep{kortemeyer2024,kortemeyer2019}.

\subsection{Epistemology, Attitudes, Motivation}
As AI transforms how scientists generate, revise, and validate knowledge, there is a pressing need to examine how AI reshapes the characteristics of the scientific enterprise and, in turn, influences perceptions of the Nature of Science \citep{CheungZhang2025}. Open questions include how generative AI alters students’ ways of knowing, the calibration of self-efficacy (especially for non-majors), and the role of metacognitive supports that help learners critique, verify, and appropriately trust AI outputs.

\subsection{Inclusivity, Accessibility, and Well-Being}
Multimodal AI offers accessibility gains (e.g., translation, captioning, alternative representations) that could broaden participation, yet differential access and potential harms (e.g., overreliance, social displacement) must be monitored \citep{salman2024,hill2025,sunday2021,herbener2025,namvarpour2025}. Rigorous studies are needed to quantify when and for whom AI improves outcomes, and to ensure that accommodations translate to durable learning.

\subsection{Teaching Techniques and Classroom Orchestration}
Generative AI can be positioned as a peer or facilitator in collaborative settings, with early work exploring impacts on Socratic dialogue, peer instruction, and group dynamics \citep{crouch2001,gregorcic2024,brookes2021}. Complementary research examines the value of deliberate \emph{tech-free} spaces to sustain sensemaking and community \citep{megowan2016,sunday2021}. The challenge is to design activity structures that harness AI’s strengths while maintaining student agency and productive discourse.

\subsection{Teacher Professional Development and AI Literacy}
Teachers need emerging forms of PCK that integrate disciplinary aims, pedagogy, and AI tool affordances/limitations. Empirical work documents factors shaping adoption decisions (trust, infrastructure, workload, policy clarity), and outlines professional learning pathways for AI literacy, prompt design, and ethical awareness \citep{nazaretsky2022,hillman2023,polverini2024b,feldmanmaggor2025}. Cross-national surveys suggest that trust relates to self-efficacy and understanding of AI, highlighting the need for calibrated conceptions that avoid both naive optimism and blanket skepticism \citep{viberg2023}.

\subsection{Ethics, Data Governance, and Infrastructures}
Ethical use of AI in education requires attention to data protection (e.g., GDPR/FERPA), environmental costs, and decolonial critiques of training corpora and deployment contexts \citep{djeki2024,zimmerle2021,dewittprat2024,jiao2024,muzanenhamo2024}. A practical frontier concerns whether non-commercial models and open infrastructures can support STEM reasoning at scale, who bears the costs of inference, and how to document and audit systems end-to-end.

\subsection{AI-Enabled Research Methodology}
AI is not only an object of study; it is increasingly a \emph{methodological partner}. Recent work explores AI-assisted analysis of large volumes of open-ended student work, from explanations to lab reports, raising questions about validity, transparency, and equity \citep{wilson2024,wan2024,chen2025,siiman2023opportunities,tansomboon2017design,wyrwich2025}. Methodological frontiers include the appropriateness of chatbots for structured interviews, automated technical transcription, and the generation of synthetic datasets for experimentation \citep{kieser2023,stanja2025}. Parallel efforts document biases introduced by pretrained embeddings and offer cautions for interpretation and deployment \citep{schleifer2024}. A live debate concerns whether non-commercial or researcher-fine-tuned LLMs can compete with commercial systems, suggesting a shift of attention from isolated models toward end-to-end \emph{systems}—data, prompts, tooling, and human oversight—optimized for educational use.

\subsection{What Is Known and What Remains Open}
Across these strands, several themes emerge. First, AI is reframing—but not replacing—the field’s longstanding commitments to epistemic cognition, equity, and methodological rigor. It extends inquiry into new domains such as multimodal reasoning, automated assessment, and synthetic data generation, while amplifying familiar concerns about validity, bias, and accessibility.

Second, many studies converge on the importance of human--AI complementarity. Evidence from curriculum design, assessment, and professional learning suggests that the most productive outcomes arise when AI augments rather than automates core educational processes. Designing for this complementarity—by clarifying where AI can enhance reasoning, feedback, and inclusion—remains a crosscutting research challenge.

Third, the rapid expansion of AI tools has outpaced conceptual and ethical frameworks for their use. Researchers increasingly call for systematic validity arguments, transparent documentation, and equitable access to open infrastructures. Ensuring interpretability and fairness in AI-supported science education is not merely a technical task but a moral and epistemic one.

Finally, open questions persist: How should science education research recalibrate its theoretical models in light of AI’s new capabilities? Which competencies—reasoning, critique, creativity—should curricula now emphasize? How can teacher education keep pace with the evolving epistemic demands of AI-augmented classrooms? Addressing these questions will require integrative designs that unite empirical insight with theory-driven reflection, setting the stage for the methodological considerations discussed in Section~\ref{sec:ways}.

\section{Ways of Inquiry in Science Education Research and How AI Is Affecting Them}\label{sec:ways}

Science education research advances through diverse ways of inquiry, including qualitative analysis, quantitative and design-based methods, and ethnography. The rapid maturation of artificial intelligence (AI) is reshaping each of these approaches, not by replacing core methodological commitments, but by expanding what counts as data, what kinds of questions are tractable, and how inferences are warranted. In this section, we synthesize what is known and outline open questions for the field.

\subsection{Quantitative Inquiry: From Parametric Assumptions to Model-Based Discovery}
Traditional quantitative approaches in science education often rely on (non)-parametric statistics with explicit assumptions about distributions, linearity, and independence. These approaches have been powerful and interpretable, but can struggle when the number of variables grows, when relationships are nonlinear, or when interactions are high-dimensional. Contemporary machine learning broadens the repertoire by handling large feature spaces and complex function classes (e.g., tree ensembles, kernel methods, deep networks), making it feasible to detect patterns and interactions that would be invisible to conventional models \citep{bishop2006,goodfellow2016}. In education studies, this capacity enables, for example, modeling the joint influence of prior knowledge, motivation, classroom discourse features, and activity structures on learning outcomes without heavy pre-specification of forms.

However, these gains come with shifts in epistemic practice. AI models may weaken reliance on distributional assumptions while introducing dependence on data curation, feature engineering, and objective functions. Validity arguments, therefore, must foreground the \emph{relationship between data and algorithm}: how training distributions, loss functions, and evaluation metrics instantiate the construct of interest; how spurious correlations are detected; and how fairness and subgroup performance are assessed. Best practice increasingly combines predictive modeling with planned confirmatory analyses and sensitivity checks, aligning model-based discovery with theory-driven inference.

\subsection{Qualitative Inquiry: Triangulation, Scale, and the ``Cyborg Analyst''}
Qualitative researchers have long emphasized iterative interpretation, triangulation across sources, and attention to context and meaning. AI expands what can be seen at what scale, particularly for audio–video records of classrooms, labs, and online learning environments. Automated or semi-automated pipelines for speech-to-text, speaker diarization, gesture and gaze detection, and event segmentation can surface candidate moments for interpretation and support principled sampling \citep{dangelo2020,krist2023,palaguachi2022,palaguachi2024}. Used reflexively, these tools become resources for triangulation rather than replacements for human judgment \citep{rosenbergkrist2021,sherin2013}.

A central challenge is maintaining an interpretivist stance while incorporating computational affordances. Analysts must ask not only \emph{what} a model detects, but \emph{how} those detections are constituted by training data, optimization targets, and error profiles. Emerging scholarship characterizes the ``cyborg analyst'': a qualitative researcher who intentionally blends human interpretive labor with AI-enabled retrieval, coding, and visualization, while openly reasoning about where algorithmic output is likely to be helpful or distorting (e.g., issues of accent, lighting, domain language, and cultural practice) \citep{dyer_krist_underreview}. We see productive use cases in \emph{assistive} coding (rapid first-pass labels for later adjudication), \emph{narrow} detection tasks (e.g., finding all graph-talk episodes), and \emph{meta-analytic} synthesis across large video corpora—always paired with human sensemaking and audit trails.

\subsection{Ethnographic and Design-Based Inquiry: New Modalities, New Responsivities}
Ethnography and design-based research (DBR) attend to activity, participation, and consequential change over time. AI adds two capabilities : first, it makes new modalities tractable at scale (continuous audio–video, interaction logs, multimodal artifacts); second, it enables \emph{responsive} interventions that adapt to learners and contexts in real time (e.g., conversational agents that scaffold argumentation, vision systems that detect representational moves). These affordances open questions about how to preserve the thick description and relational ethics central to ethnography when data volume grows dramatically, and how to document AI-mediated design rationales and their local contingencies within DBR cycles \citep{Badran2025_AIAugNetnography}.

\subsection{What Is Known and What Remains Open}
Across methods, several lessons are consolidating. First, AI extends data \emph{volume} and \emph{variety}, but interpretability, construct validity, and equity require deliberate methodological design. Second, human and machine inferences proceed via different processes; leveraging their complementarity—rather than substituting one for the other—yields more robust findings. Third, AI can productively serve as an instrument for triangulation throughout the interpretive process \citep{rosenbergkrist2021,sherin2013}. Open questions include: What additional \emph{thinking practices} should science education scholars cultivate (e.g., systematic ``blurring'' between human and algorithmic perspectives)? How should we conceptualize effective collaboration with AI tools (division of labor, provenance, accountability)? Which data types or research questions are now feasible that previously were not, and what ethical infrastructures are required? Finally, how can we most effectively mobilize the reliability and scale of computational tools to \emph{augment} (rather than overshadow) theory-driven inquiry?

\section{Navigating Complexity and Uncertainty When Integrating AI into Science Education Research}
Integrating AI into science education (research) introduces layered complexity and genuine uncertainty: many stakeholders, diverse research aims, and rapid technical shifts create interdependencies and feedback loops that simple checklists miss. To support that research in this space remains generative, adaptable, and responsible, we advocate for adopting a systems thinking approach \cite{herdliska2024}. Systems thinking enables researchers to recognize interdependencies, feedback loops, and unintended consequences across different layers of educational practice and research. 

\subsection{A Heuristic for Navigating Complexity}
We propose a heuristic to support science education researchers in navigating this evolving space. The heuristic is operationalized through a table that juxtaposes stakeholder groups (students, educators, parents, researchers, society, funders) against a set of guiding dimensions: goals, potential uses of AI, research contributions, emergent problems, limitations, and associated ethical, legal, or cost-related challenges. By systematically considering each cell, researchers are prompted to broaden their aperture, identifying overlooked stakeholders, anticipating unintended consequences, and clarifying the ways their research contributes to both practice and knowledge.

\begin{table}[ht]
\centering
\caption{Stakeholder-by-dimension heuristic table}
\footnotesize
\setlength{\tabcolsep}{4pt}
\renewcommand{\arraystretch}{1.5}
\begin{tabularx}{\textwidth}{|l| *{6}{>{\centering\arraybackslash}X|}}
\hline
& \textbf{Goals}
& \textbf{Potential uses of AI}
& \textbf{Research contributions}
& \textbf{Emergent problems}
& \textbf{Limitations}
& \textbf{Challenges (ethical, legal, cost)} \\
\hline
\textbf{Students}    & & & & & & \\ \hline
\textbf{Educators}   & & & & & & \\ \hline
\textbf{Parents}     & & & & & & \\ \hline
\textbf{Researchers} & & & & & & \\ \hline
\textbf{Society}     & & & & & & \\ \hline
\textbf{Funders}     & & & & & & \\ \hline
\end{tabularx}
\end{table}

The process of using the table unfolds in several steps:

1. Define the research question or problem space.  

2. Identify primary stakeholder goals and explore ripple effects for secondary stakeholders.  

3. Consider the type of AI engagement (learning with AI, learning about AI, or research using AI).  

4. Map advances the research that can be made and the potential gaps it can fill.  

5. Anticipate risks—threats to validity, ethical issues, accessibility, costs, or legal constraints.  

6. Document and report the methodological and ethical standards followed.

This process emphasizes reflexivity: researchers must account for how AI tools shape not only their findings but also the conditions under which those findings become meaningful.

\subsubsection{Use Case: A K–12 Data Analysis Tool}
Consider the case of a researcher designing an AI-integrated tool that allows students to issue natural language commands to manipulate datasets. Students can say, for example, “show me only the plants with diseased leaves,” and the tool translates this request into a sequence of data wrangling steps. The immediate research question is how such a tool helps students pursue their investigative questions in science.

Students’ goals include successfully completing assignments, earning satisfactory grades, and engaging meaningfully with data. Teachers, however, face the goal of supporting students in these inquiries and aligning instruction with curricular standards, while parents may seek to understand and encourage their children’s progress. Each group interacts with AI differently: students learn with AI, teachers teach with it, and parents support learning through it. By situating the research question within the heuristic, the researcher recognizes new needs, such as professional learning opportunities for teachers and transparent communication with parents.

The potential contributions of this research extend beyond immediate stakeholders. Other researchers, tool designers, and curriculum developers may use these findings to refine educational technologies. Yet new challenges also arise: Should students still learn “manual” data-cleaning skills without AI? How sustainable is the integration of such tools beyond the project context? Addressing these concerns may involve embedding explanatory feedback into the tool, ensuring that students not only benefit from automation but also understand the underlying processes.

\subsubsection{Use Case: Scaling Responsive Teaching in Higher Education}
At the university level, AI presents both opportunities and dilemmas for responsive teaching. Adaptive AI systems promise to scale individualized feedback, but they also risk amplifying inequities if access, transparency, or validity are compromised. Using the heuristic, stakeholders can be mapped: Students value timely feedback and fairness; instructors aim to maintain pedagogical agency; administrators weigh costs and compliance; and researchers must ensure methodological rigor. Applying the framework highlights trade-offs: the promise of personalization is coupled with environmental costs, ethical concerns about surveillance, and challenges in maintaining human interpretive oversight \citep{rosenberg2021,krist2023,palaguachi2024}.

\subsection{Summary}
This section argued that integrating AI into science education research entails coupled technical, pedagogical, and ethical dependencies. We introduced a systems-thinking heuristic—operationalized via a stakeholder-by-dimension table—to make these dependencies explicit and to support reflexive, rigorous design. The K–12 and higher education use cases illustrate how the heuristic helps align aims, anticipate side effects, and select proportional safeguards. The framework thus offers a shared language and a bridge from complexity to action; next, we translate these insights into concrete implications for training, infrastructure, and standards.

\section{Implications and Recommendations for the Science Education Research Community}

AI is not just another topic in science education (research); it reconfigures our questions, methods, and evidence base. Building on Sections 2–5—which outlined what AI is changing in science and classrooms, how it intersects core strands of our field, how it reshapes ways of inquiry, and why a systems-thinking lens is needed—this section translates those insights into actionable guidance. We organize the implications around three cross-cutting enables that determine what the community can credibly know and do: training, infrastructure, and standards. We then offer guidance on partnering with EdTech and industry, acknowledging that many AI tools used in education originate outside academia. Throughout, the recommendations aim to (a) safeguard epistemic vigilance and equity, (b) leverage human–AI complementarity, and (c) embed transparency and compliance as routine practice (“standards-by-design”). The goal is pragmatic: to help researchers, departments, professional societies, funders, and policy makers convert the preceding analyses into durable capabilities and near-term steps. 

\subsection{Training: Capacities for a Moving Target}

For (early-career) researchers, the AI landscape can feel like a moving target. A durable preparation includes (a) \textbf{conceptual foundations} in statistics and machine learning, (b) \textbf{architectural literacy} (e.g., transformers for LLMs) and their strengths/limits for educational data, and (c) \textbf{computational fluency} to translate ideas into working code using mainstream toolchains and MLOps (short for Machine Learning Operations, the practice of automating and streamlining the development, deployment, and maintenance of machine learning models in production) practices. In practice, methodological choice points (classic inference vs.\ predictive modeling; small, task-specific models vs.\ general-purpose LLMs) should be tied to construct definitions and data properties rather than novelty alone. Considering the scale of what such a durable preperation entails, however, the question of collaboration also becomes more pressing. Is it really feasible that science education researchers develop these capacities or do we need to develop more interdisciplinary collaborations?

Equally important are \textbf{ethical foundations}. Researchers must be able to reason about privacy and data governance (e.g., FERPA/GDPR-aligned workflows), energy and water usage in model training/inference, and bias and fairness from pretraining data through deployment \citep{yang_beil_2024,urquieta_dib_2024,masley_2025,loukina_2019,feng_2023}. We recommend lab- or center-level \emph{communities of practice} that periodically surface case-based dilemmas (e.g., sampling bias in log data; water use of cloud regions; subgroup performance audits) and document decisions publicly as part of open methods.

Finally, we advise a pragmatic stance toward capability building: \textbf{avoid competing with frontier model labs}. Rather than investing scarce time in training bespoke foundation models, build \emph{replaceable} pipelines around robust APIs or mid-sized open models such as Mistral that can be hosted locally, with evaluation harnesses that allow rapid swap-in of newer back-ends. When specialized performance is needed (e.g., short-answer scoring aligned to rubrics), smaller or similarity-based models may outperform generic LLMs, provided appropriate data and evaluation are in place \citep{horbach_zesch_2019,ferreira_mello_2025,bexte_2022,bexte_2023,bexte_2025}.

\subsection{Infrastructure: The Driver of What We Can Know}

Infrastructure is not merely a backdrop; it actively drives the field’s epistemic possibilities. We distinguish four intertwined layers:

\paragraph{Data infrastructure.}
Open-access datasets and FAIR-aligned standards (Findable, Accessible, Interoperable, Reusable) allow cumulative progress and replication while respecting legal/ethical constraints \citep{furlough_2010,bowers_choi_2023,yang_beil_2024}. Common schemas for anonymization, consent tracking, and documentation make cross-site synthesis feasible.

\paragraph{Computing infrastructure.}
Access to GPUs/TPUs and cloud resources remains uneven, yet national programs (e.g., ACCESS in the U.S.) and institutional HPC centers mitigate barriers \citep{wang_2020,boerner_2023}. On the software side, shared pipelines that encode responsible-AI checkpoints (provenance, bias checks, privacy protections) help labs move fast without breaking ethics \citep{vyhmeister_2023}.

\paragraph{Human infrastructure.}
Interdisciplinarity is mandatory: science education, learning sciences, human-computer-interaction, and Natural Language Processing / Machine Learning each contribute essential expertise. Professional learning for \emph{AI literacy for researchers}—from prompt/pipeline design to bias auditing—should be institutionalized, alongside research–practice partnerships that provide authentic contexts and two-way value \citep{yang_2025,getenet_2019,kozma_2000,popova_2024,shah_2024}.

\paragraph{Scholarly infrastructure.}
The field’s evaluation and dissemination pathways should reflect AI’s velocity. Grant programs and review cycles can privilege \emph{research aims and infrastructure} over brittle, tool-specific deliverables; tenure and promotion should recognize conference proceedings, preprints, datasets, code, and registered reports where appropriate \citep{pane_2014,roschelle_2017}. Journals and societies can adopt fast-track tracks for methods/tooling papers with strong open-science artifacts.

\subsection{Standards: Meeting (and Evolving) the Bar}

Standards for validity, reliability, ethics, and security remain non-negotiable, but the \emph{ways} we meet them need updating. We highlight five categories:

\begin{itemize}
  \item \textbf{Human oversight and epistemic vigilance.} Treat AI as an instrument whose output is always situated; require documented human adjudication for consequential judgments.
  \item \textbf{Design standards.} For interventions using AI, pre-specify contrasts (e.g., human-only vs.\ AI-augmented feedback), guard against data leakage, and register outcomes/analyses where feasible.
  \item \textbf{Ethical, security, and privacy standards.} Align with FERPA/GDPR; externalize threat models for re-identification; disclose environmental and monetary costs where material \citep{yang_beil_2024,vyhmeister_2023}.
  \item \textbf{Validity and reliability with AI in the loop.} When using automated scoring or analytics, report construct mapping, calibration procedures, subgroup analyses, and error decompositions \citep{zesch_2023,horbach_zesch_2019,wang_rubrics_2019,gombert_2023}.
  \item \textbf{Fairness and bias.} Use multiple, sometimes competing, fairness definitions; combine data balancing and augmentation with in-/post-processing mitigations; complement metrics with qualitative checks \citep{loukina_2019,ganin_2016,zhao_2018,feng_2023}.
\end{itemize}

Where appropriate, we encourage \emph{standards-by-design}: embed checklists and audit hooks into data and MLOps pipelines so that compliance is a byproduct of routine practice rather than an after-the-fact patch.

\subsection{EdTech and Industry: Partnering for Impact}

Educational Technology (EdTech) platforms and research-intensive companies can be strategic partners when collaborations are principled. Evidence from intelligent tutoring and learning-at-scale work suggests that effectiveness hinges on curricular alignment, teacher integration, and school-level implementation—not just algorithms \citep{vanlehn_2011,kulik_fletcher_2016,pane_2014,roschelle_2017,blanc_2023}. From a computer science perspective, mapping educational tasks to precise ML problem types prevents overuse of generative LLMs where smaller or more stable models suffice \citep{laarmann_quante_2022,zesch_2023,ferreira_mello_2025}. Partnerships should formalize \emph{data-sharing agreements, evaluation plans, and openness commitments} (artifacts, benchmarks, and reproducible code) while recognizing that many standard ML benchmarks underrepresent school populations and may suffer data leakage; domain-sensitive evaluation is essential.

\section{Recommendations by stakeholders}

Based on the previous points, we here highlight the most urgent actions needed to ensure that the integration of artificial intelligence (AI) into science education and science education research is both responsible and generative by stakeholder.

\subsection{Funding Bodies}
\begin{itemize}
    \item \textbf{Support faster, higher-risk research.} In the age of rapid AI development, funding mechanisms should allow for shorter review cycles and riskier exploratory projects to keep pace with technological change. 
    \item \textbf{Invest in infrastructure.} Prioritize open datasets, access to computing resources, and shared pipelines with embedded ethical safeguards to ensure equity and cumulative progress. 
    \item \textbf{Establish stable, scalable, and continuous (24/7) inference infrastructure for education.}Without such infrastructure, educational AI models cannot be effectively deployed or accessed, regardless of how advanced or open-weight they are. Inference—though often considered non-research—has distinct requirements from research computing and is essential for ensuring equitable access to AI systems developed through educational research.
    \item \textbf{Require ethical and environmental accountability.} Include privacy, bias, and sustainability reporting as part of grant deliverables.
\end{itemize}

\subsection{Educational Policy}
\begin{itemize}
    \item \textbf{Integrate AI responsibly into curricula.} Teach AI concepts within science disciplines, rather than as isolated content. 
    \item \textbf{Redesign assessment frameworks.} Emphasize reasoning, modeling, and explanation over outcomes that can be easily automated. 
    \item \textbf{Ensure equity of access.} Address disparities in teacher training and infrastructure across schools and regions.
\end{itemize}

\subsection{Professional Organizations (e.g., NARST)}
\begin{itemize}
    \item \textbf{Accelerate publishing and knowledge exchange.} Develop alternative, faster forms of dissemination (e.g., preprints, methods notes, shared benchmarks) to keep pace with AI’s rapid evolution. 
    \item \textbf{Build shared infrastructure.} Coordinate repositories for datasets, code, and evaluation tools to make research more cumulative and reproducible. 
    \item \textbf{Develop ethical and methodological standards.} Provide the field with common benchmarks for validity, fairness, and responsible AI use.
\end{itemize}

\subsection{Science Education Departments}
\begin{itemize}
    \item \textbf{Embed AI literacy in teacher education.} Prepare future teachers to critically and productively use AI in science classrooms. 
    \item \textbf{Reframe researcher training.} Equip graduate students with both traditional methods and AI-augmented approaches, emphasizing epistemic vigilance. 
    \item \textbf{Foster interdisciplinary collaboration.} Build bridges to computer science, ethics, and HCI to support responsible and innovative use of AI.
\end{itemize}

\paragraph{Bottom line.}
Investing in people, platforms, and policies—in that order—will determine whether AI accelerates rigorous, equitable science education research. With the right training, infrastructures, and standards, our community can move fast \emph{and} build things that last.

\bibliography{references}

\end{document}